\documentclass[seceq]{ptptex}

\usepackage{graphicx}
\title{%        %You can use \\ for explicit line-break
Effective theory of color superconductivity%
}

\author{%       %Use \scshape  for the family name
Deog Ki \textsc{Hong}\footnote{ e-mail address: dkhong@pusan.ac.kr}
}

\inst{%         %Affiliation, neglected when [addenda] or [errata]
Department of Physics, Pusan National University,
Busan, 609-735, South Korea\\
}

\abst{%         %this abstract is neglected when [addenda] or [errata]
We briefly review an effective theory of QCD at high baryon density,
describing the relevant modes
near the Fermi surface.  The high density effective theory
has properties of  reparametrization invariance and
gauge invariance, maintained in a subtle way.
It also has a positive measure, allowing lattice simulations at high baryon density.
We then apply it to gapless
superconductors and discuss recent proposals to resolve the magnetic instability
of gapless superconductivity.}

\begin{document}

\maketitle

%\section{Title page}
%Please be aware that there are two fields for the
%title and authors.
%One for the {\bf full authors' name}\\
%\hspace*{1cm} Daisuke Jido and Atsushi Hosaka,\\
%and the other with initials\\
%\hspace*{1cm} D. Jido and A. Hosaka.\\
%Similarly, one {\bf for full title}\\
%\hspace*{1cm} Recognition of Shapes and Colors at YKIS06 While Thinking\\
%and the other for {\bf short one}\\
%\hspace*{1cm} Recognition of Shapes and Colors

\section{Introduction}
Quantum Chromodynamics (QCD) is now the accepted theory of strong interactions,
consistent with all experimental data.
One of the salient features of QCD is that strong interactions become weaker and weaker
at high energy or at short distances due to the chromomagnetic interaction of gluons,
a phenomenon called ``asymptotic freedom". The prediction of QCD on how
the coupling depends on the scale has been well tested by numerous experiments. %,
%including the $e^+-e^-$ scattering at LEP, heavy quark spectra, and lattice calculations.
The logical consequence of the asymptotic freedom is then
that hadronic matter, bound by the strong interactions,
must undergo a phase transition to quark matter,
when the hadronic matter gets extremely squeezed or heated up,
at the critical temperature
or the critical chemical potential  being of order of $\Lambda_{\rm QCD}$,
the characteristic scale of QCD.

The phase transition of QCD at finite temperature but at zero density has been confirmed
by lattice calculations to occur at temperature about $200~{\rm MeV}$~\cite{Karsch:2000hh}.
The lattice calculation of QCD at finite baryon density, however, has not made much progress, since
it suffers from the notorious sign problem.
The Euclidean partition function of QCD at finite density, used for lattice simulations,
does not have a positive-definite measure, because the determinant of the Dirac
operator, $M=\gamma_E^{\mu}D_E^{\mu}+\mu\gamma_E^4$,
is no longer positive definite, when the chemical potential, $\mu$, is present.

There have been several attempts to overcome
this difficulty at finite but small density by several methods~\cite{Fodor:2001au,Hands:2007by}.
At high density it was recently pointed out that the sign problem is
due to the fast modes, quarks with energy much larger than the chemical potential, suggesting
that the sign problem is either mild or absent for physics near the Fermi surface~\cite{Hong:2002nn}.

While the lattice simulation struggles to overcome the sign problem,
there have been much progress in understanding QCD at high baryon density
(or dense QCD for short) from the
theoretical side~\cite{Rajagopal:2000wf}, using effective theories or models of dense QCD.
The ground state of dense QCD is believed to be a color superconducting
state, where quarks near the Fermi surface form Cooper pairs. Unlike the electron
superconductor, where the pairing is mediated by phonons, the Cooper-pairing in quark matter
is due to the gluon exchange interaction, since quarks in antisymmetric in color are attractive
to each other.
Though the quark-antiquark channel is most attractive,
it is costly to form quark-antiquark condensate since quarks in Dirac sea has to be excited
above the Fermi sea.

Recently, it is found that the color superconducting quark matter
exhibits rather rich phases because quarks have several flavors with different masses,
interacting electroweakly~\cite{Ruster:2005jc}.
One interesting phase is the so-called gapless color superconducting state~\cite{Shovkovy:2003uu},
where the Fermi
sea is partially filled, though quarks do pair, allowing gapless excitation of quarks.
However, it is soon pointed out that the gapless color superconductor has imaginary Meissner
mass, indicating its instability~\cite{Huang:2004bg}.
In this talk we discuss the high density effective theory of dense QCD
and apply it to study the gapless color superconductivity and propose two resolutions
of the instability of the gapless superconductor~\cite{Hong:2005jv,Gorbar:2005rx}.

\section{High Density Effective Theory}

The relevant degrees of freedom of dense QCD at low energy are quarks and holes near the
Fermi surface together with screened gluons, since quarks deep in the Fermi sea and also antiquarks
are decoupled from the low-energy dynamics due to Pauli blocking.
The effective theory of quarks and holes in
dense QCD is derived in references~\cite{Hong:1998tn,Nardulli:2002ma}.

Since the quark chemical potential in dense QCD is very large,
$\mu\gg\Lambda_{\rm QCD}$, the velocity of quarks and holes
is conserved under a typical QCD interaction and thus
we may decompose the momentum of quarks near the Fermi surface
\begin{equation}
p^{\mu}=\mu\,v^{\mu}+l^{\mu},
\end{equation}
where the residual momentum $\left|l^{\mu}\right|<\mu$ and $v^{\mu}=(0,\vec v_F)$ with
Fermi velocity $\vec v_F$.
The effective Lagrangian of quarks and holes is given as
\begin{equation}
\label{treeL} {\cal L}_{\rm Q}=
Z_{\shortparallel}\,\bar\psi_+i\gamma_{\shortparallel}\cdot D\,\psi_+-Z_{\perp}\,
\bar\psi_+
\gamma^0\frac{(\gamma_{\perp}\cdot D\,)^2}{2\mu}\psi_+ ~+~ \cdots,
\end{equation}
where $D=\partial-ig_s\,A$ is the covariant derivative and
$\gamma_{\shortparallel}^{\mu}=(\gamma^0,\hat v_F\,\hat v_F\cdot\vec \gamma)
=\gamma^{\mu}-\gamma^{\mu}_{\perp}$,
\begin{equation}
\psi_+(\vec v_F,x)=\frac{1+\vec\alpha\cdot \hat v_F}{2}e^{-i\mu\vec v_F\cdot\vec x}\psi(x),
\end{equation}
with $\vec\alpha=\gamma^0\vec \gamma$ and $\hat v_F=\vec v_F/\left|\vec v_F\right|$.
We note that the effective theory enjoys a reparametrization invariance,
\begin{equation}
\vec v_F\mapsto \vec v_F+\frac{\delta \vec l_{\perp}}{\mu},\quad
\vec l\mapsto \vec l-\delta l,
\end{equation}
under which the Lagrangian~(\ref{treeL}) is invariant. The reparametrization invariance
is quite useful when we  calculate the loop corrections, since it restricts
the form of counter terms. For instance the reparametrization invariance
requires  $Z_{\shortparallel}=Z_{\perp}$.

\begin{figure}[tbh]
\centering
\includegraphics[width=4.0cm]{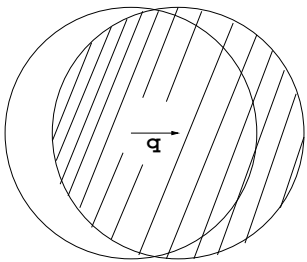}
\caption{Spectral Flow. The circle denotes a Fermi surface.}\label{fig2}
\end{figure}

The Lagrangian for gluons contains in addition to the usual kinetic term
the terms coming from the modes far away from the Fermi surface,
\begin{equation}
{\cal L}_{\rm G}=-\frac{1}{4g_s^2}F_{\mu\nu}^2-\frac{1}{2}M^2A_{\mu}A_{\nu}\,
g^{\mu\nu}_{\perp}+\cdots,
\label{gluon}
\end{equation}
where the Debye screening mass $M^2=N_fg_s^2\mu^2/(2\pi^2)$ for $N_f$
flavors.~\footnote{The ellipsis denotes terms with higher powers of gauge fields,
which are known as hard-dense-loop Lagrangian~\cite{Schafer:2006yf}.}
The high density effective theory of dense QCD is then given as
\begin{equation}
{\cal L}_{\rm HDET}={\cal L}_{\rm Q}+{\cal L}_{\rm G}\,.
\end{equation}
It is important that one should include the Debye screening mass and also higher
order terms in (\ref{gluon}) to maintain the gauge invariance, since the Fermi
surface is not gauge invariant and $\psi_+$ modes alone are not enough
to maintain the gauge-invariance. Under
a gauge transformation, $U(x) = e^{i\vec q\cdot\vec x}$,
the energy level shifts and so does the Fermi surface (see Fig.~\ref{fig2})
%figure---------------------------------------------
%figure---------------------------------------------
\begin{equation}
E=\vec l\cdot\vec v_F\mapsto E=\vec l\cdot \vec v_F+\vec
q\cdot\vec v_F.
\end{equation}

Since the mode decomposition of quark fields into modes near the Fermi surface
and modes far away
from the Fermi surface is not gauge invariant, the effect of modes away from
the Fermi surface should be added in the effective Lagrangian
otherwise the gauge-invariance
will be lost. Let's consider for instance the current-current correlation function,
%\begin{equation}
\begin{eqnarray}
\Pi^{\mu\nu}(p)=\int_xe^{ip\cdot x}
\left<J^{\mu}(x)J^{\nu}(0)\right>=
-{iM^2}\!\int{{\rm d}\Omega_{\vec v_F}\over 4\pi}
\left({-2\vec p\cdot\vec v_FV^{\mu}V^{\nu}\over p\cdot V
+i\epsilon \,\vec p\cdot\vec v_F}+g_{\perp}^{\mu\nu}\right),\label{correlation}
\end{eqnarray}
%\end{equation}
where $\epsilon\to0^+$. We see that the Debye mass term or $g_{\perp}^{\mu\nu}$
in the parenthesis in Eq.~(\ref{correlation})
is essential for the current conservation and thus for the gauge invariance,
\begin{equation}
p_{\mu}\Pi^{\mu\nu}(p)=0\,.
\end{equation}
\section{Positivity at asymptotic density}
For quarks very near the Fermi surface ($E\ll\mu$) the Fermi surface is almost flat
and one can show that there is a symmetry that pairs the eigenvalues of the Dirac operator:
\begin{equation}
M_{\rm eft}=\gamma^{E}_{\parallel}\cdot D(A)=
\gamma_5\,M_{\rm eft}^{\dagger}\,\gamma_5\,.
\end{equation}
Since the eigenvalues of the Dirac operator is pure imaginary, $i\lambda$
($\lambda$ being real),
the determinant of the Dirac operator is positive semi-definite:
\begin{equation}
\det\, M_{\rm eft}=\det\, M_{\rm eft}^{\dagger}=\left(\prod_{\lambda}\lambda^2\right)^{1/2}\,.
\end{equation}
After integrating out the fast modes, the partition function for dense QCD becomes
\begin{equation}
Z(\mu)=\int {\rm d}A~\det M_{\rm eff}(A)e^{-S_{\rm eff}(A)},
\end{equation}
and the effective action is given as
\begin{eqnarray}
S_{\rm eff}(A)=\int{\rm
d}^4x_E\left({1\over4}F_{\mu\nu}^aF_{\mu\nu}^a +{M^2\over
16\pi}\sum_{\,\vec v_F}A_{\perp\mu}^{a}A_{\perp\mu}^{a}\right)
+\cdots,
\label{eff_action}
\end{eqnarray}
where the ellipsis denotes terms with higher powers of the gauge fields and
also the terms coming from the higher order terms in the quark Lagrangian~(\ref{treeL}).

We see that the measure of the partition function is positive definite if we do
not keep the higher order terms in the effective action, allowing us to simulate on a lattice.
One byproduct of the positivity at asymptotic density is that one can use
the Vafa-Witten theorem~\cite{Vafa:1983tf} to prove
the Color-Flavor-Locked phase~\cite{Alford:1998mk} is the true ground
state at very high baryon density\cite{Hong:2002nn},
since the vector symmetry can not be  spontaneously broken
when the measure is positive.

The size of the higher order terms can be estimated, using the naive dimensional analysis.
We find the size of corrections to the leading terms to be at a scale $\Lambda\ll\mu$
%\begin{equation}
$\frac{\alpha_s}{2\pi}\,\frac{\Lambda}{\mu}\,$.
%\end{equation}
Therefore the sign problem of dense QCD is mild or absent if $\alpha_s\,\Lambda\ll2\pi\mu\,$.

\section{Application: New phases in gapless superfluids}
The color superconductivity has a rich phase structure due to stress on pairing quarks.
Since quarks in dense matter
pair with different flavors for spin-zero pairing, the flavor-dependent mass and
the electroweak interaction among different flavors leads to the difference
in the chemical potential, $2\delta\mu$, between pairing quarks.

\begin{figure}[tbh]
\centering
\includegraphics[width=7.0cm]{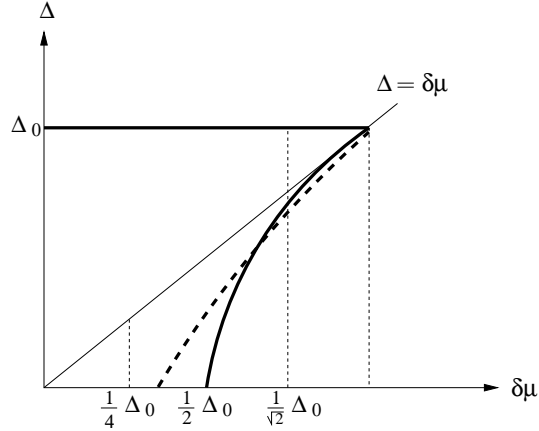}
\caption{Solutions to the gap equation and
the charge neutrality equation.}\label{fig3}
\end{figure}

If we solve the Cooper-pair gap equation
for dense matter under stress,
\begin{equation}
\frac{\partial\Omega}{\partial\Delta}=0\,,
\end{equation}
we find two branches of solutions (See Fig.~\ref{fig3}): When $\Delta>\delta\mu$,
$\Delta=\Delta_0$, the BCS phase, and when $\Delta<\delta\mu$,
$\Delta=\sqrt{\Delta_0\left(2\delta\mu-\Delta_0\right)}$, the Sarma phase~\cite{Sarma}.
%figure---------------------------------------------
%figure---------------------------------------------
The BCS phase is unstable for $\delta\mu>\Delta_0/\sqrt{2}$, known as the Clogston
instability~\cite{clogston}:
When the stress is big enough, the Cooper pairs break. On the other hand,
the Sarma phase is unstable for all values of $\delta\mu$ in
$\frac12\Delta_0<\delta\mu<\Delta_0$, for which the solution exists.

Recently it was pointed out that the Sarma phase might be stable if one imposes the charge
neutrality condition on quark matter~\cite{Shovkovy:2003uu},
\begin{equation}
\frac{\partial\Omega}{\partial\mu_Q}=0\,,
\end{equation}
where $\mu_Q$ is the electric charge chemical potential.
(The charge neutrality has to be satisfied for quark matter inside of compact stars,
otherwise too much energy will build up at the surface of quark matter, destabilizing
whole system.) The free energy difference between the neutral Sarma phase and the neutral
quark matter is found to be in the mean field approximation~\cite{hong}
\begin{equation}
\left.\delta\Omega\right|_{Q=0}=\left.\Omega_{\rm Sarma}\right|_{Q=0}
-\left.\Omega_{\rm free}\right|_{Q=0}=
-\frac{{\bar\mu}^2}{2\pi^2}\,\left(\delta\mu_0\right)^2\,
\left[1-\left(\frac{\Delta_0}{2\delta\mu_0}-1\right)^2\right]\,,
\end{equation}
where $\bar\mu$ is the average chemical potential and $\delta\mu_0$ is half of the
chemical potential difference of neutral, unpaired quark matter. The charge neutrality
condition stabilizes the Sarma phase if
\begin{equation}
\frac{\Delta_0}{4}\le\delta\mu_0\le\frac{\Delta_0}{\sqrt{2}}\,.
\end{equation}

\begin{figure}[tbh]
\centering
\includegraphics[width=7.0cm]{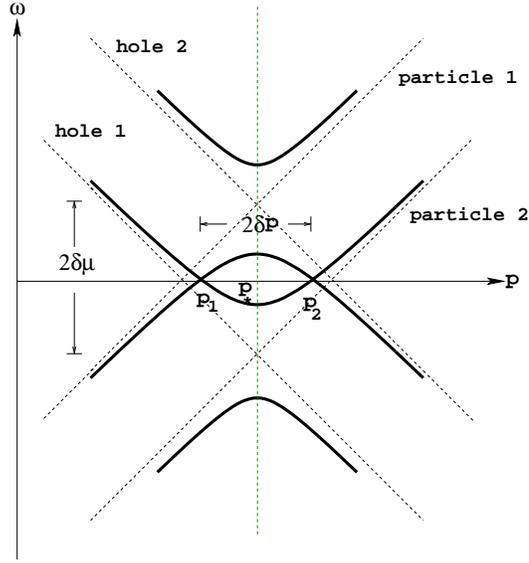}
\caption{The energy dispersion relation of quasi quarks and holes.
The Fermi sea is partially filled, having inner and outer Fermi surfaces
at $p_1$ and $p_2$.}\label{fig4}
\end{figure}

The energy dispersion relation of quarks in the Sarma phase is given as
\begin{equation}
\omega(\vec
p)=\pm\left(\delta\mu\pm\sqrt{\epsilon^2(\vec p)+\Delta^2}\right)\,,
\end{equation}
which is plotted in Fig.~\ref{fig4}.
%figure---------------------------------------------
%figure---------------------------------------------
At low energy the modes near the Fermi surface are relevant. The modes near the Fermi
surface are gapless and have either quadratic or linear dispersion relations, depending
on the strength of the stress:
\begin{eqnarray}
\omega(\vec p\,)\simeq \left\{\begin{array}{ll}
{\eta\,\vec v_i\cdot\vec l}\,,\,\, &{\rm
if}\,\,
\left|\vec v_i\cdot\vec l\,\right|<\delta p\\
{\frac{(\vec v_i\cdot\vec
l)^2}{2\delta\mu}}\,,\, &{\rm if}\,\,
\delta p<\left|\vec v_i\cdot\vec l\,\right|<\delta\mu\\
\vec v_i\cdot\vec l\,,\,\,& {\rm if}
\,\,\bar\mu\gg\left|\vec v_i\cdot \vec l\,\right|>\delta\mu\,,
\end{array}
\right.
\end{eqnarray}
where $\vec v_{1}$ and $\vec v_2$ are the Fermi velocities of quarks near the Fermi
surfaces, $p_1$ and $p_2$, respectively.

When $\delta \mu\approx\Delta$ or $\delta p\approx0$, the gapless modes are quadratic all
the way down to the Fermi surface at $p_*\simeq p_1\simeq p_2$.
After integrating out the irrelevant degrees of freedom,
we get an effective Lagrangian for quadratic gapless modes,
\begin{eqnarray}
{\cal L}_{\rm eff}= \sum_{\vec
v_*}\Psi^{\dagger}(\vec v_*,x) \left[i\partial_t+\frac{(\vec
v_*\cdot\vec\nabla)^2}{2\,\delta\mu}\right] \Psi(\vec
v_*,x)+\frac{\kappa}{2}\left(\Psi^{\dagger}\Psi\right)^2+\cdots\,.
\end{eqnarray}
Under the scaling $E\mapsto s\,E$, the action
$S=\int {\rm d}t {\rm d^3l}\,{\cal L}_{\rm eff}$ is invariant and we have
\begin{equation}
\vec l_{\shortparallel}\mapsto s^{1/2}\,\vec
l_{\shortparallel},\quad\psi_{\vec{ v_*}}(t,\vec{\, l}\,)\mapsto
s^{-1/4} \psi_{\vec{ v_*}}(t,\vec{\, l}\,)\,.
\end{equation}
The four Fermi-interaction is relevant for incoming fermions
with opposite momenta~\footnote{If not opposite, the interaction is marginal.},
regardless of the sign,
\begin{equation}
\kappa\mapsto s^{-1/2}\,\kappa\,.
\end{equation}
The four-Fermi interaction generated by the quantum effects of
the irrelevant modes will open a secondary gap at the Fermi surface, which is only
power-suppressed in couplings~\cite{Hong:2005jv,Alford:2005kj},
\begin{eqnarray}
\Delta_s\simeq6.85
\,\kappa^2\,\left(\frac{\nu_*}{v_*}\right)^2\delta\mu
\,.
\end{eqnarray}

When the stress is large enough
or
the infrared cutoff for the quadratic modes
\begin{equation}
\Lambda^{\rm IR}_{\rm
quadratic}=\frac{1}{2\delta\mu}\left(\delta\mu^2-\Delta^2\right)>\Delta_s\,,
\end{equation}
the secondary gap does not open, because the RG
running is not long enough.
%Instead, the condensate of Goldstone currents
%develops.
The system is then dominated by linearly gapless modes at
$\omega<\Lambda^{\rm IR}_{\rm
quadratic}$.

Integrating out all the modes except the linearly gapless modes, we get
an effective potential for the gauge fields with
$%\begin{equation}
V_1^{\mu}=(\eta\bar q+\delta
q,\eta\bar q\,\vec v_1)$ and $V_2^{\mu}=(\eta\bar q-\delta
q,\eta\bar q\,\vec v_2)$,
\begin{eqnarray}
V(A)&=&\frac{1}{2}m_M^2{\vec A\,}^2
-\frac{\eta}{3}\left(\nu_1v_1^2+\nu_2v_2^2\right)e^2{\bar
q\,}^2\,{\vec A\,}^2
\left[\ln\left(\frac{M^2}{{\vec A\,}^2}\right)+3\right]\\
&&
-\frac{1}{2}m_D^2A_0^2-\frac{1}{\eta}\left(\nu_1+\nu_2\right)\,
\left(\eta\bar q+\delta q\right)^2e^2{A_0}^2
\left[\ln\left(\frac{M^2}{{A_0}^2}\right)+3\right]\,, \label{e_pot}
\end{eqnarray}
where $\nu_1$ and $\nu_2$ are the area of the inner and outer Fermi surfaces,
respectively, and the electric charges of pairing quarks are
$q_1=\bar q+\delta q$ and $q_2=\bar q-\delta q$,
upon imposing the  renormalization conditions,
\begin{eqnarray}
\left.\frac{\partial^2V}{\partial
A_0^2}\right|_{{A_0}=M}=-m_D^2\,,\quad
\frac{1}{3}\,\delta_{ij}\left. \frac{\partial^2V}{\partial
A_i\partial A_j}\right|_{{\vec A\,}^2=M^2} =m_M^2\,. \label{r_cond}
\end{eqnarray}
The minimum of the potential occurs at
\begin{eqnarray} \left<\left({\vec A-\vec
\nabla\varphi\,}\right)^2\right> \simeq{\delta\mu}^2\,
\exp\left[2-\frac{4\,\nu_*\,v_*^2}
{\eta\left(\nu_1v_1^2+\nu_2v_2^2\right)} \right]\,.
\end{eqnarray}
We find that gapless superfluids are stabilized by spontaneously generating
Nambu-Goldstone currents. One can further  show that the Meissner mass for this phase
is nonnegative but directional~\cite{Hong:2005jv}.

%\vfill\eject

\section{Conclusions}
We have discussed an effective theory of dense QCD, called high density
effective theory. The theory has a reparametrization invariance and the gauge invariance
is maintained by counter terms. Furthermore the theory has a positive semi-definite measure
at the leading order. Lattice simulation should be possible for high density quark matter.
By the positivity at high density one can establish a rigorous theorem like the Vafa-Witten
theorem.

As an application of the effective theory, we study the gapless color superconductivity,
which occurs when the color superconductors are under stress.
When the stress is comparable to the gap, the gapless modes have a quadratic
dispersion relation and open a secondary gap at the Fermi Surface, stabilizing the system.
The secondary gap
is only power suppressed in couplings due to peculiar scaling properties of the quadratic
gapless modes.
When the stress is much larger than the gap, the relevant modes are linearly gapless modes.
By calculating the  Coleman-Weinberg potential for the gauge fields, we show that
the gapless superfluids become stabilized by spontaneously generating
Nambu-Goldstone currents.
%\eject
\section*{Acknowledgements}
The author thanks the organizers of YKIS2006 on ``New Frontiers in QCD"
for the invitation and the hospitality during his stay at the
Yukawa Institute for Theoretical Physics at Kyoto University.
The work of D.~K.~H. was supported for two years
by Pusan National University Research Grant.

%%%%%%%%%%%%%%%%%%%%%%%%%%%%%%%%%%%%%%%%%%%%%%%%%%%%%%%%%%%%%
% Some macros are available for the bibliography:
%  o for general use
%    \JL : general journals                 \andvol : Vol (Year) Page
%  o for individual journal
%    \AJ   : Astrophys. J.           \NC         : Nuovo Cim.
%    \ANN  : Ann. of Phys.           \NPA, \NPB  : Nucl. Phys. [A,B]
%    \CMP  : Commun. Math. Phys.     \PLA, \PLB  : Phys. Lett. [A,B]
%    \IJMP : Int. J. Mod. Phys.      \PRA - \PRE : Phys. Rev. [A-E]
%    \JHEP : J. High Energy Phys.    \PRL        : Phys. Rev. Lett.
%    \JMP  : J. Math. Phys.          \PRP        : Phys. Rep.
%    \JP   : J. of Phys.             \PTP        : Prog. Theor. Phys.
%    \JPSJ : J. Phys. Soc. Jpn.      \PTPS       : Prog. Theor. Phys. Suppl.
% Usage:
%  \PRD{45,1990,345}          ==> Phys.~Rev.\ \textbf{D45} (1990), 345
%  \JL{Nature,418,2002,123}   ==> Nature \textbf{418} (2002), 123
%  \andvol{B123,1995,1020}    ==> \textbf{B123} (1995), 1020
%%%%%%%%%%%%%%%%%%%%%%%%%%%%%%%%%%%%%%%%%%%%%%%%%%%%%%%%%%%%%

\end{document}